\documentclass[aps,prl,nofootinbib,nobibnotes,frontmatterverbose,reprint,11pt,twocolumn,letterpaper,preprintnumbers,showkeys,floatfix,amsmath,amssymb,dvipsnames,svgnames]{revtex4-2}

\usepackage{amsmath}
\usepackage{mathptmx}
\usepackage{eurosym}
\usepackage{graphicx}
\usepackage{braket}
\usepackage{booktabs}
\usepackage{placeins}
\usepackage{academicons}
\usepackage{svg}
\usepackage{dcolumn}
\usepackage{multirow}
\usepackage{amssymb}
 \usepackage{subfigure}
  \usepackage{caption}
\usepackage{bbm}

\usepackage{amsfonts}
\usepackage[left=1.5cm,right=1.1cm,top=1.5cm,bottom=1.5cm]{geometry}
\usepackage[hypertexnames=false]{hyperref}
\usepackage{tikz}
\usepackage{orcidlink}
\usepackage{cleveref}   % Load cleveref second

\definecolor{darkgreen}{rgb}{0.1,0.6,0.1}
\definecolor{darkblue}{rgb}{0.2,0.2,0.7}
\definecolor{darkred}{rgb}{0.7,0,0}

\definecolor{light gray}{RGB}{220,220,220}
\definecolor{dark purple}{RGB}{108,0,217}
\definecolor{pink}{RGB}{190,20,100}
\definecolor{orang}{RGB}{193,63,0}
\definecolor{green}{RGB}{11,98,17}
\definecolor{darkpink}{RGB}{153,0,76}
\definecolor{bluegreen}{RGB}{0,102,102}
\definecolor{greenlagan}{RGB}{0,102,0}
\definecolor{redgreen}{RGB}{102,102,0}
\definecolor{Redgreen}{RGB}{153,76,0}
\definecolor{vividviolet}{rgb}{0.62, 0.0, 1.0}
\definecolor{amaranth}{rgb}{0.9, 0.17, 0.31}
\definecolor{palatinateblue}{rgb}{0.15, 0.23, 0.89}
\definecolor{brightpink}{rgb}{1.0, 0.0, 0.5}
\definecolor{cornflowerblue}{rgb}{0.39, 0.58, 0.93}
\definecolor{deepcarminepink}{rgb}{0.94, 0.19, 0.22}
\definecolor{radicalred}{rgb}{1.0, 0.21, 0.37}
 
\definecolor{beamer@PRD}{RGB}{46,48,146}
\hypersetup{
     breaklinks=true,
    pdfstartview={FitH},    % fits the width of the page to the window
    colorlinks=true,       % false: boxed links; true: colored links
    linkcolor=darkblue,          % color of internal links
    citecolor=darkblue,        % color of links to bibliography
    filecolor=darkblue,      % color of file links
    urlcolor=darkblue,           % color of external links
    anchorcolor=cyan,      % Color for anchor text
    linktocpage=true
}

\begin{document}
	{\vskip .1cm}
% \date{\today}
\newcommand\be{\begin{equation}}
\newcommand\ee{\end{equation}}
\newcommand\bea{\begin{eqnarray}}
\newcommand\eea{\end{eqnarray}}
\newcommand\bseq{\begin{subequations}} %solo con amsmath
\newcommand\eseq{\end{subequations}}
\newcommand\bcas{\begin{cases}}
\newcommand\ecas{\end{cases}}
\newcommand{\p}{\partial}
\newcommand{\f}{\frac}
\renewcommand{\thefootnote}{\fnsymbol{footnote}}
% \renewcommand{\thesubfigure}{\alph{subfigure}}
% Define \email and \orcid manually

\title{\Large  Saturation of Chaos Bound in a Phenomenological Rotating Black Hole–Effective Matter System at Low-Temperature Limit\normalsize}

\author {Reza Pourkhodabakhshi \orcidlink{0000-0003-4375-4972}   }
\email[Contact author : ]{reza.pk.bakhshi@gmail.com}
\affiliation{ Departament de F\' \i sica Cu\' antica i Astrof\'\i sica and Institut de Ci\`encies del Cosmos, 
Universitat de Barcelona, Mart\'i Franqu\`es, 1, 08028
Barcelona, Spain.}
\affiliation{Department of Physics and Physical Oceanography, Memorial University of Newfoundland, St. John’s, Newfoundland and Labrador A1B 3X7, Canada}

 %-------------------------------------------------------------------------------------------------------------------------------------

\begin{abstract}
We study the chaotic behavior of a phenomenological system of rotating black holes (BHs) surrounded by radiation, dust, and dark matter as effective matter fluids. We show that the chaos bound is saturated --but never violated-- when the density parameter \( k \rightarrow 1 \), which corresponds to a high dark matter concentration in the halo. This saturation is independent of radiation and dust density, occurring in the low-temperature regime where the BH's surface gravity and Hawking temperature decrease. Our results suggest that rotating BHs subjected to a phenomenological representation of a more realistic matter environment can scramble information as efficiently as allowed by physical laws in this limit. This work links quantum chaos research with phenomenological BH-matter configurations, providing a motivation for studies of chaos in astrophysical BHs.

\vspace{1em} % Space between the abstract and the DOI
\noindent

 \end{abstract}

\maketitle

\textit{Introduction}— 
In classical mechanics, chaos refers to the exponential divergence of trajectories that begin from nearby initial conditions, with this sensitivity quantified by a positive Lyapunov exponent. In quantum systems, analogous behavior is described by the out-of-time-ordered correlator (OTOC),
\begin{equation}
C(t) = -\langle [W(t), V(0)]^2 \rangle_\beta,
\end{equation}
where $V$ and $W$ are Hermitian operators, and the thermal expectation value is defined as $\langle . . .  \rangle_\beta = \mathrm{Tr}(e^{-\beta H} \cdot)/Z$, evaluated at inverse temperature $\beta = 1/T$. At early times, $C(t)$ typically grows as $C(t) \sim e^{2\lambda_L t}$, allowing $\lambda_L$ to serve as the quantum analog of the classical Lyapunov exponent. This interpretation holds in the regime where the dissipation time $t_d \sim \beta$ and the scrambling time 
$ t_s \sim \lambda_L^{-1} \log(1/  \hbar ) $
are well separated \cite{1969JETP...28.1200L}.

Maldacena, Shenker, and Stanford proposed a universal upper bound on $\lambda_L$ in thermal quantum systems, based on holographic duality and the behavior of shockwaves near black hole horizons in AdS spacetime \cite{Maldacena_2016,Shenker_2014,Shenker2_2014}. This bound has been intensively studied in the Sachdev-Ye-Kitaev (SYK) model \cite{Sachdev_1993,Polchinski_2016}, a disordered one-dimensional fermionic system with random interactions \cite{Maldacena22_2016}, which also possesses a gravitational dual \cite{Kitaev_2018}. 

The Lyapunov exponent has also been widely investigated in black hole backgrounds, particularly for probe particles in curved spacetimes \cite{ref20,ref21,ref22,ref23,ref24,ref25,ref26,ref27,ref28,ref29,ref30,ref31,ref32,ref33,ref34,ref35,ref36} (see also \cite{ref37,ref38,ref39,ref40,ref41,ref42,ref43,ref44,ref45,ref46,ref47}), and through the AdS/CFT correspondence \cite{ref48,ref49,ref50,ref51,ref52,ref53,ref54,ref55}. In this framework, a probe particle near a black hole horizon is dual to a boundary field theory operator. Ref. \cite{ref56} examined the Lyapunov bound for such particles in static, spherically symmetric black holes (BH) under external forces like electric and scalar fields. Around the local maximum of the effective potential—modeled by an inverted harmonic potential, which induces chaos \cite{ref50,ref57,ref58,ref59}—they found that the maximal Lyapunov exponent saturates the Maldacena bound.

However, with higher-spin external interactions, the bound was shown to be violated \cite{ref56}. In a broader survey, Ref. \cite{ref60} computed Lyapunov exponents for various BH including asymptotically flat, AdS, and de Sitter (dS) Reissner–Nordström (RN) solutions. For dS BH, they found that the bound holds only near the event horizon, while being violated when the potential maximum lies further out \cite{ref60}.

Astrophysical BHs are not isolated from matter \cite{Haardt2016}, 
%and may be enclosed within a profile of it, 
therefore it is important to study the phenomenological effect of matter around the $\mathrm{BH}$ on the chaos as well. 
While presently it is not  known what kind of matter dominates the region around the $\mathrm{BH}$, one can parametrize
different types of matter by using the Kiselev spacetime \cite{Kiselev_2003}. This describes a family of exact solutions in which the rotating $\mathrm{BH}$ is surrounded by three types of anisotropic fluid matter; radiation, dust, dark matter with Equation-of-States (EoSs) $\alpha=1 / 3,0$ and $-1 / 3$, respectively \cite{Cuadros_Melgar_2021}.

Understanding the chaotic dynamics of Kiselev, a phenomenological BH-matter system, offers insight into their connection to quantum description within the AdS/CFT framework and SYK models.

\textit{Setup for BH–Effective Matter System}— 
Starting from a spherically symmetric solution and employing the standard Newman-Janis algorithm~\cite{osti_4608715}, along with the refinements proposed in~\cite{Azreg_A_nou_2014}, one can obtain the rotating counterpart of the charged Kiselev black hole. In Boyer-Lindquist coordinates \((t, r, \theta, \varphi)\), the resulting metric takes the form presented in~\cite{Rahaman_2010}.

\begin{equation}
\begin{aligned}
&ds^2 = -\left(\frac{\Delta_k - a^2 \sin^2 \theta}{\Sigma}\right) dt^2 + \frac{\Sigma}{\Delta_k} dr^2\\
&- 2a \sin^2 \theta \left(\frac{\Delta_k - (r^2 + a^2)}{\Sigma}\right) dtd\varphi + \Sigma d\theta^2  \\
&+ \sin^2 \theta \left(\frac{(r^2 + a^2)^2 - a^2 \Delta_k \sin^2 \theta}{\Sigma}\right) d\varphi^2,
\end{aligned}
\label{eq:Kiselev-Metric-QNM}
\end{equation}
where
\begin{equation}
\Sigma = r^2 + a^2 \cos^2 \theta, \quad \Delta_k = r^2 - 2mr + a^2 + Q^2 - kr^{1 - 3\alpha}.
\label{eq:delta-Kiselev}
\end{equation}
Here, \( m \) and \( a \) denote the mass and spin parameter of the black hole, respectively. The constant \( k \) arises as an integration parameter that quantifies the contribution of the surrounding matter field, effectively distinguishing this solution from the standard Kerr-Newman metric, which is recovered when \( k = 0 \). This constant is related to a dimensionless parameter \( \mathcal{J} \) by the expression \( k = \mathcal{J} m^{1 + 3\alpha} \). The parameter \( \alpha \) represents the equation of state (EoS) and characterizes the dominant form of matter surrounding the black hole. In phenomenologically more realistic scenarios, multiple matter components may coexist. However, we demonstrate that radiation and dust contributions can be absorbed into the effective mass \( M \) and charge \( Q \), respectively. This formulation provides a useful framework for analyzing the chaotic dynamics influenced by different matter components, including radiation (\( \alpha = \frac{1}{3} \)), dust (\( \alpha = 0 \)), and dark matter (\( \alpha = -\frac{1}{3} \)).

 %In this paper, we limit our investigation to three types of fluid matter around the BH: radiation $(\alpha = \frac{1}{3})$, dust $(\alpha = 0)$, and dark matter $(\alpha = -\frac{1}{3})$. 

\bigskip

It is important to note that dark matter can be modeled in several ways; in this work, it is treated as a fluid component of the matter field~\cite{Rahaman_2010,Xu_2018}.  

Additionally, two key clarifications should be made regarding the metric in Eq.~\eqref{eq:Kiselev-Metric-QNM}. First, contrary to some interpretations in the literature, this metric does not correspond to a perfect fluid configuration, but instead describes an anisotropic fluid. For an in-depth discussion and critique of this issue, the reader is referred to the work by Visser~\cite{Visser_2020}.

For \eqref{eq:Kiselev-Metric-QNM}, the tt-component of stress-energy tensor is given by

\begin{equation}
    \begin{aligned}
        &T_{tt} = 
        \resizebox{!}{18pt}{[}  
         8 Q^2 r^{1 + 6\alpha} \left(3 a^2 + 2 (Q^2 - 2 m r + r^2) - 
a^2 \cos(2 \theta) \right) \alpha \\
        & + 48 k^2 r^3  +  k r^{3 \alpha} \resizebox{!}{15pt}{(} 
-16 Q^2 r^2 (1 + 3 \alpha) \\
&+ 
3 \alpha \left(16 (2 m - r) r^3 + a^4 (1 - 3 \alpha) - 
4 a^2 r^2 (5 + 3 \alpha) \right) \\
&+ 
12 a^2 r^2 \alpha (1 + 3 \alpha) \cos(2 \theta) + 
3 a^4 \alpha (-1 + 3 \alpha) \cos(4 \theta)
 \resizebox{!}{16pt}{)} \resizebox{!}{18pt}{]}  \\
        &\times \left( 2 r^{6 \alpha +1} \left(a^2 \cos (2 \theta )+a^2+2 r^2\right)^3 \right)^{-1}.
    \end{aligned}
    \label{eq:T00-rotating}
\end{equation}
 For  $\alpha=\frac{1}{3}, \ -\frac{1}{3},\  0$, we obtain
 \begin{equation}
 \begin{aligned}
     & T^{\textbf{Rad}}_{tt} = \frac{
4 (k - Q^2) \left(2 k -3 a^2 - 2 Q^2 + 4 m r - 2 r^2 + 
a^2 \cos(2 \theta)\right)
}{
\left(a^2 + 2 r^2 + a^2 \cos(2 \theta)\right)^3
} ,\\
     & T^{\textbf{DM}}_{tt} = \left[
 -a^4 k + 4 a^2 (3 Q^2 + 2 k r^2) - 
 4 a^2 Q^2 \cos(2 \theta) \right. \\
&+ a^4 k \cos(4 \theta)   \left.+ 
 8 (Q^2 + k r^2)(Q^2 + r (-2 m + r - k r))  \right] \\
&
\times \left(
(a^2 + 2 r^2 + a^2 \cos(2 \theta))^3 \right)^{-1}
 ,\\
     & T^{\textbf{Dust}}_{tt} = \frac{
4 Q^2 \left(3 a^2 + 2 \left(Q^2 - (k + 2 m) r + r^2\right) - 
a^2 \cos(2 \theta)\right)
}{
\left(a^2 + 2 r^2 + a^2 \cos(2 \theta)\right)^3
} ,
     \end{aligned}
     \label{eq:T00-alpha-cases}
 \end{equation}
The term \( T^{\textbf{Dust}}_{tt} \big|_{r \sim \infty} = \frac{Q^2}{r^4} + O\left[\left(\frac{1}{r}\right)^5\right] \) reflects the fact that, in the case of dust (\( \alpha = 0 \)), the spacetime reduces to a vacuum solution of general relativity. This corresponds to the Kerr-Newman geometry, modified by a shift in the mass term \( 2m \to 2m + k \). This special case serves as a reference for contrasting the influence of dark matter with that of standard dust-like matter.

The asymptotic form of \( T_{tt} \) from Eq.~\eqref{eq:T00-alpha-cases} in the limit \( r \to \infty \) is given by:

 \begin{equation}
     \begin{aligned}
         & T^{\textbf{Rad}}_{tt}|_{r\sim \infty}=\frac{Q^2 -k}{r^4}+O\left[\left(\frac{1}{r}\right)^5\right], \\
         & T^{\textbf{DM}}_{tt}|_{r\sim \infty} = \frac{k(1-k) }{r^2}+O\left[\left(\frac{1}{r}\right)^3\right].
     \end{aligned}
     \label{eq:T00-asymptotic}
 \end{equation}
For the case \( \alpha = -\frac{1}{3} \), the parameter must be restricted to \( k < 1 \), as values \( k \geq 1 \) lead to an incorrect metric signature at spatial infinity. From Eq.~\eqref{eq:T00-asymptotic}, the weak energy condition \( T_{tt} \geq 0 \) holds if \( k^\textbf{Rad} \leq 0 \) (with \( Q^2 < |k| \), noting that \( Q \) is typically negligible in astrophysical scenarios) and \( k^\textbf{DM} \geq 0 \). These constraints on \( k \), as confirmed by Eq.~\eqref{eq:T00-alpha-cases}, ensure that the weak energy condition is satisfied not only asymptotically but throughout the spacetime exterior to the event horizon.

Specifically, for \( \alpha = -\frac{1}{3} \), one finds \( \rho = \frac{T_{tt}}{-g_{tt}} \big|_{r \to \infty} = \frac{k}{r^2} \) and \( P_{rr} = \frac{T_{rr}}{g_{rr}} \big|_{r \to \infty} = -\frac{k}{r^2} \), indicating that \( \rho = -P_{rr} \sim 1/r^2 \). This \( r^{-2} \) decay behavior is a hallmark of dark matter contributions to the energy-momentum tensor, consistent with models of galactic halo dynamics~\cite{Kiselev_2003,Matos_2000}. It is important to note that in this scenario, the spacetime is not asymptotically flat, as the metric does not approach Minkowski form at large \( r \). Nonetheless, this geometry remains valid in the halo region, where dark matter is presumed to dominate.

In contrast, for \( \alpha = +\frac{1}{3} \), the stress-energy tensor is traceless, \( \operatorname{tr}(T_{ij}) = 0 \), reflecting the conformal invariance typical of radiation fields governed by Maxwell theory. This is consistent with the fact that the corresponding metric reduces to the Kerr-Newman solution.

In summary, while dust alters only the black hole mass and radiation acts as an effective charge, it is dark matter that introduces qualitatively new spacetime features. Therefore, analyzing the \( \alpha = -\frac{1}{3} \) case in the context of rotating charged Kiselev BH provides insight into the collective effects of all three halo components.

%%%%%%%%%%%%%%%%%%%%%%%%%%%%%%%%%%%%%
%%%%%%%%%%%%%%%%%%%%%%%%%%%%%%%%%%%%%%

\textit{Chaos in Phase Space}—  
In classical dynamical systems, Lyapunov exponents quantify the sensitivity to initial conditions by analyzing perturbations around a reference trajectory in phase space. In black hole spacetimes, photon rings exhibit characteristic Lyapunov behavior, which serves as a diagnostic of chaotic dynamics. In this work, we adopt the method developed in~\cite{Kan_2022}.
Our analysis begins with a Polyakov-type action describing the motion of a test particle in this geometry.
\begin{equation}
\begin{aligned}\label{eq:action2}
S &= \int \mathcal{L}_p ds =\int d s\Bigg[\frac{1}{2 \hat{A}} \Bigg(-\frac{1}{\Sigma}\left(\Delta_k-a^2 \sin^2 \theta\right)+\frac{\Sigma}{\Delta_k} \dot{r}^2\\
&+\Sigma \dot{\theta}^2+\frac{2 a \sin^2 \theta}{\Sigma}\left(\Delta_k-\left(r^2+a^2\right)\right) \dot{\varphi} + \frac{a q_{\text{tp}} Q \sin^2\theta}{\Sigma} \dot{\varphi}  \quad \\
&+\frac{\sin^2 \theta}{\Sigma}\left(\left(r^2+a^2\right)^2-\Delta_k a^2 \sin^2 \theta\right) \dot{\varphi}^2-\frac{\hat{A}}{2} m_{tp}^2 - \frac{q_{\text{tp}} Q r}{\Sigma} \Bigg)   \Bigg] ,
\end{aligned}    
\end{equation}
where $\hat{A}$ is an auxiliary field and $m_{tp}$ is the mass of the test particle and we worked on a static gauge $X_0=s$. We focus on equatorial plane $\theta=\pi / 2$.
The action is invariant under translation for $\varphi$. In order to obtain an effective action, we first need to calculate the angular momentum,

\begin{equation}
\begin{aligned}\label{eq:angular_momentum}
L_{\varphi} &=\frac{a}{\hat{A} r^2}\left(\Delta_k-\left(r^2+a^2\right)\right)+\frac{1}{\hat{A} r^2}\left(\left(r^2+a^2\right)^2-\Delta_k a^2\right) \dot{\varphi}\\
& + \frac{a q_{\text{tp}}  Q}{r} .
\end{aligned}    
\end{equation}

To erase the auxiliary field $\hat{A}$ from the Lagrangian, we use the equation of motion for $\hat{A}$,

\begin{equation}
\begin{aligned}\label{eq:eom_A}
&-\hat{A}^2 m_{tp}^2=
-\frac{1}{r^2}\left(\Delta_k-a^2\right)+\frac{2 a}{r^2}\left(\Delta_k-\left(r^2+a^2\right)\right) \dot{\varphi}\\
&+\frac{1}{r^2}\left(\left(r^2+a^2\right)^2-\Delta_k a^2\right) \dot{\varphi}^2+\frac{r^2}{\Delta_k} \dot{r}^2 .
\end{aligned}    
\end{equation}

Using \eqref{eq:angular_momentum} and \eqref{eq:eom_A}, we define an effective Lagrangian as,
\begin{equation}
\begin{aligned}\label{eq:lagrangian_eff}
\mathcal{L}_{\text{eff}} &= \mathcal{L}_p -L_{\varphi} \dot{\varphi} = \hat{A}\left[-\frac{m_{tp}^2}{2}-\frac{1}{2} \frac{(a q_{\text{tp}} Q - L_\varphi r)^2}{\left(r^2 + a^2\right)^2 - \Delta_k a^2}\right] \\
&- \frac{1}{2 \hat{A}}\left[\frac{\Delta_k - a^2}{r^2} - \frac{r^2}{\Delta_k} + \frac{a^2\left(\Delta_k - \left(r^2 + a^2\right)\right)^2}{r^2\left(\left(r^2 + a^2\right)^2 - \Delta_k a^2\right)}\right] \\
& - \frac{a(a q_{\text{tp}} Q - L_\varphi r)\left(\Delta_k - \left(r^2 + a^2\right)\right)}{r\left(\left(r^2 + a^2\right)^2 - \Delta_k a^2\right)} - \frac{q_{\text{tp}} Q}{r},
\end{aligned}    
\end{equation}

where

\begin{align}\label{eq:A_solution}
\hat{A} = r \sqrt{\frac{\Delta_k^2-\left(\left(r^2+a^2\right)^2-\Delta_k a^2\right) \dot{r}^2}{\gamma(r)}},
\end{align}

with $\gamma(r)$ defined as

\begin{align}\label{eq:alpha_r}
\gamma(r) = \Delta_k m_{tp}^2\bigg[\left(\left(r^2+a^2\right)^2-\Delta_k a^2\right) +(aq_{\text{tp}}Q-L_\varphi)^2\bigg].
\end{align}

For $r > r_{+}$, the function $\gamma(r)$ remains positive. Our analysis centers on the particle’s motion near the local maximum of a potential. In this region, the particle's initial velocity is relatively small. Consequently, its motion in the $r$-direction can be described using the non-relativistic approximation, assuming $\dot{r} \ll 1$. Under this condition, we derive the effective Lagrangian for the non-relativistic particle from
\eqref{eq:lagrangian_eff} as

\begin{align}\label{eq:lagrangian_nr}
\mathcal{L}_{\mathrm{eff}} = \frac{1}{2} \mathcal{K}(r) \dot{r}^2-\mathcal{V}_{\mathrm{eff}}(r)+O\left(\dot{r}^4\right) ,
\end{align}

where

\begin{align}\label{eq:K_r}
\mathcal{K}(r) = \frac{r \sqrt{\gamma(r)}}{\Delta_k^2}
\end{align}
 
 and

\begin{equation}
\begin{aligned}
&\mathcal{V}_{\mathrm{eff}}(r) = \frac{1}{\left(r^2+a^2\right)^2-\Delta_k a^2} \Bigg( r \sqrt{\gamma(r)} \\
& \quad - a L_\varphi \Delta_k + \left(r^2+a^2\right) a L_\varphi \Bigg) .
\label{eq:V_eff}    
\end{aligned}
\end{equation}

To analyze the particle's motion around the local maximum, we determine the position of the local extrema $r_0$ by solving the equation $\mathcal{V}_{\text{eff}}^{\prime}(r) = 0$. Expanding the effective Lagrangian \eqref{eq:lagrangian_nr} around $r = r_0$ with a small perturbation $r(s) = r_0 + \epsilon(s)$, we obtain  

\begin{equation}
\mathcal{L}_{\mathrm{eff}} = \frac{1}{2} \mathcal{K}\left(r_0\right) \left( \dot{\epsilon}^2 + \lambda^2 \epsilon^2 \right) ,
\label{eq:L_eff_expanded}
\end{equation}

where we omit constant and higher-order terms. In particular, the coefficient of $\epsilon^2$ defines the Lyapunov exponent

\begin{equation}
\lambda^2 = -\frac{\mathcal{V}_{\mathrm{eff}}^{\prime \prime}\left(r_0\right)}{\mathcal{K}\left(r_0\right)} .
\label{eq:Lyapunov}
\end{equation}

\begin{figure*}[t]
    \centering
    % First row
    \begin{subfigure}{ }
        \centering
        \begin{minipage}{0.3\textwidth}
            \centering
            \includegraphics[width=\textwidth]{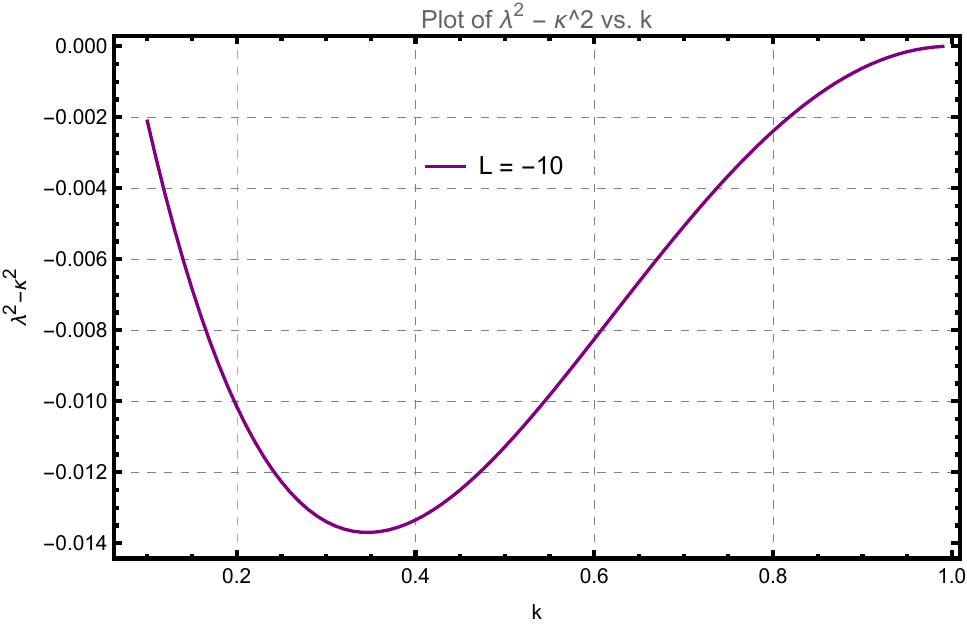}
            \caption*{(a)$\lambda^2 - \kappa^2$ vs $k$ for $L=-10$.}
        \end{minipage}
        \hspace{15pt}
        \begin{minipage}{0.3\textwidth}
            \centering
            \includegraphics[width=\textwidth]{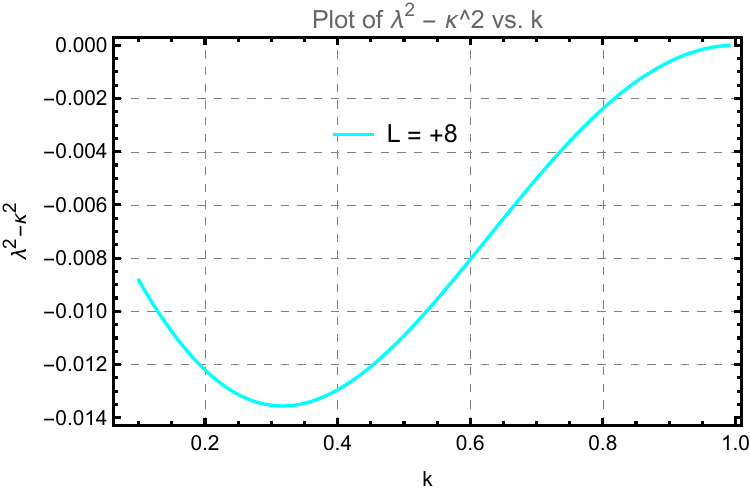}
            \caption*{(b) $\lambda^2 - \kappa^2$ vs $k$ for $L\varphi= +8$.}
        \end{minipage}
                \hspace{15pt}
        \begin{minipage}{0.3\textwidth}
            \centering
            \includegraphics[width=\textwidth]{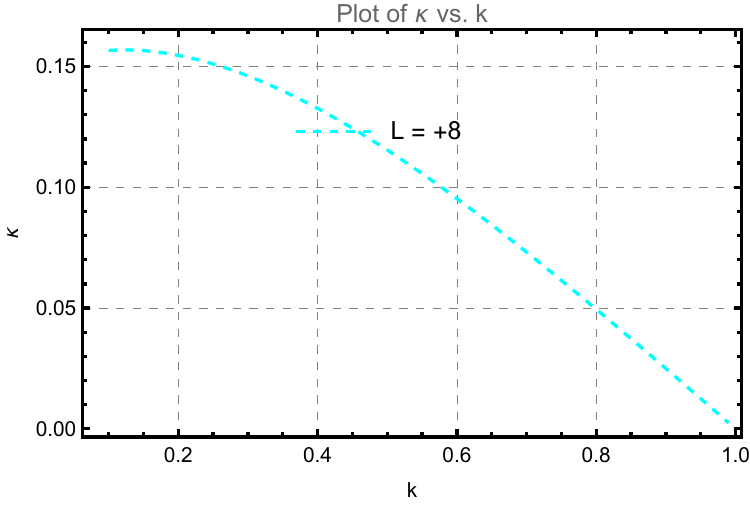}
            \caption*{(c) $\kappa$ vs $k$ for $L_\varphi= +8$.}
        \end{minipage}
        \caption{(a), (b) Numerical analysis of chaos $\lambda^2 - \kappa^2$ in an effective BH-matter system for parameter $k$ evolving from $0.1$ to $0.99$ for two different choices of $L=-10 m$ and $+8m$ from left to right, respectively.
        (c)Numerical analysis of surface gravity $\kappa$ in an effective BH-matter system for parameter $k$ evolving from $0.1$ to $0.99$ for the choice of $L=+8m$.
        We fixed  $a=0.9 m$,  $m_{\text{tp}}=0.01m$, $Q=0.1m$ and   $q=0.02m$. ($m=1$ ). Observer is equatorial, $\theta_0=\frac{\pi}{2}$. Note here with $\alpha=-\frac{1}{3}$, $k\equiv \mathcal{J}$ is dimensionless. }
        \label{fig:k-plots-2}
    \end{subfigure}
% Comment to prevent errors
    % Second row
    \begin{subfigure}{ }
        \centering
        \begin{minipage}{0.3\textwidth}
            \centering
            \includegraphics[width=\textwidth]{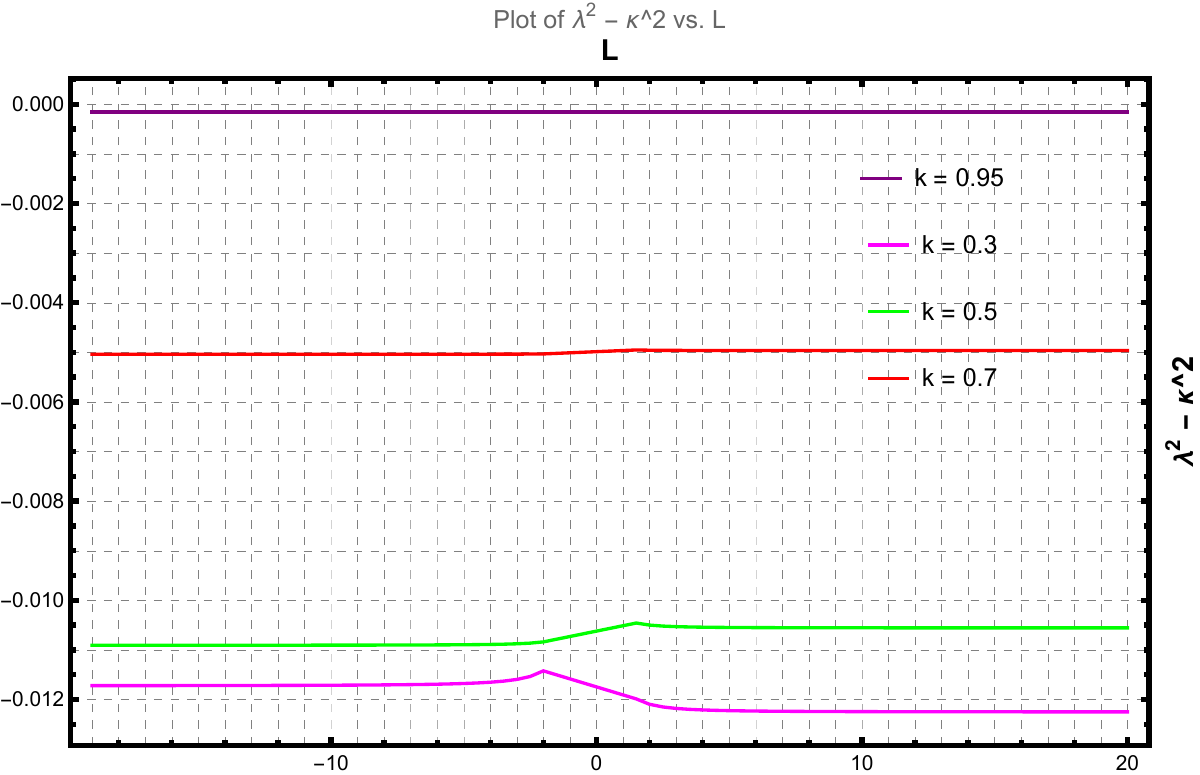}
            \caption*{(a) $\lambda^2 - \kappa^2$ vs $L_\varphi$ for different choices of $k$.}
        \end{minipage}
        \hspace{15pt}
        \begin{minipage}{0.3\textwidth}
            \centering
            \includegraphics[width=\linewidth]{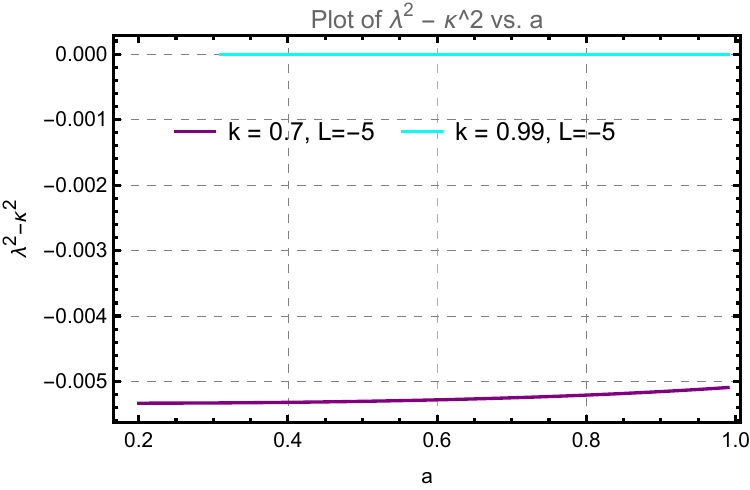}
            \caption*{(b) $\lambda^2 - \kappa^2$ vs $a$ for different choices of $k$}
        \end{minipage}
        \hspace{15pt}
        \begin{minipage}{0.3\textwidth}
            \centering
            \includegraphics[width=\linewidth]{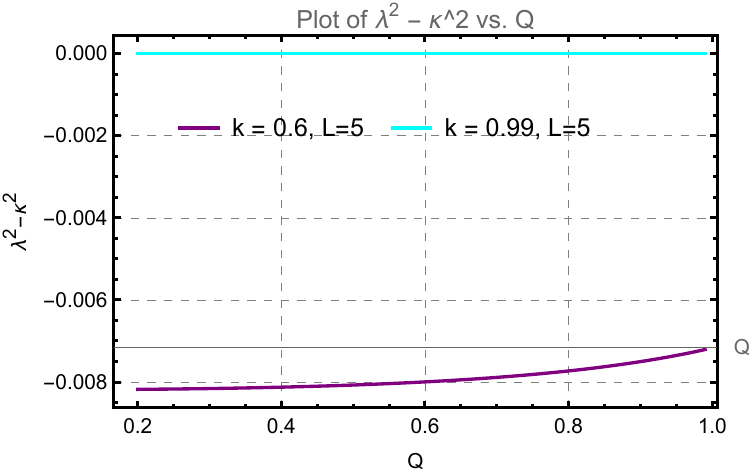}
            \caption*{(c) $\lambda^2 - \kappa^2$ vs $Q$ for different choices of $k$}
        \end{minipage}
        \caption{(a) Numerical analysis of chaos $\lambda^2 - \kappa^2$ in an effective BH-matter system for parameter $L_\varphi$ evolving from $-20$ to $+20$ for four different choices of $k=0.3$,$0.5$, $0.7$ and $0.95$. We set $a=0.95 m$ and $Q=0.1 m$ . 
        (b)  Numerical analysis of chaos $\lambda^2 - \kappa^2$ in an effective BH-matter system for parameter $a$ evolving from $0.2m$ to $1.0 m$ for two different choices of $k=0.7$ and $0.99$. We set $L_\varphi=-5$ and $Q=0.1m$.
        (c) Numerical analysis of chaos $\lambda^2 - \kappa^2$ in an effective BH-matter system for parameter $Q$ evolving from $0.2m$ to $1.0 m$ for two different choices of $k=0.6$ and $0.99$. We set $L_\varphi=+5$ and $a=0.85 m$.
        In all cases, we fixed $m_{\text{tp}}=0.01m$ and   $q=0.02m$. ($m=1$ ). Observer is equatorial, $\theta_0=\frac{\pi}{2}$. Note here with $\alpha=-\frac{1}{3}$, $k\equiv \mathcal{J}$ is dimensionless. }
        \label{fig:L-plots-2}
    \end{subfigure}
% Comment to prevent errors
  % Comment to prevent errors
\end{figure*}

This exponent  measures the
sensitivity to the initial condition. In quantum systems, the Lyapunov exponent $\lambda$ derived from OTOCs satisfies the bound
$
\lambda \leq \frac{2\pi T}{\hbar},
$
with $T$ the system temperature. For BH, using the Hawking temperature $T_{\text{BH}}$, this becomes
$\lambda \leq \kappa,$
where $\kappa$ is the surface gravity $\kappa = \frac{r_+ - r_-}{2(r_+^2 +a^2)}$ (Horizons $r_\pm $ are roots of $\Delta_k$). For chaotic motion, this implies
$
\lambda^2 \leq \kappa^2.
$
We apply this to a probe particle near a charged Kiselev BH, where the radial motion yields an effective inverse harmonic potential, setting the maximal $\lambda$.

\textit{Numerical Treatment}— 
In the rotating charged Kiselev geometry, characterized by \( \Delta_k \) as defined in Eq.~\eqref{eq:delta-Kiselev}, the local extrema of the effective potential \( \mathcal{V}_{\text{eff}}(r) \) are determined by solving \( \mathcal{V}^\prime_{\text{eff}}(r) = 0 \) resulting \( r = r_0 \). This condition is solved numerically, and the result is used in evaluating the Lyapunov exponent via Eq.~\eqref{eq:Lyapunov}. Our numerical analysis explores the behavior of \( \lambda^2 - \kappa^2 \) under variations of the parameters \( k \), \( Q \), \( L_\varphi \), and \( a \).

As previously discussed, it is sufficient to focus on the case \( \alpha = -\frac{1}{3} \) to capture the influence of a dark matter halo, since dust (\( \alpha = 0 \)) and radiation (\( \alpha = \frac{1}{3} \)) contribute to the geometry effectively as modifications to the black hole mass and charge, respectively.

The limits \( k = 1 \) and \( k = 0 \) correspond to a metric singularity and the Kerr-Newman solution, respectively. Thus, we restrict our analysis to the interval \( 0 < k < 1 \), which encapsulates the dark matter contribution. By varying \( Q \), we also account for the influence of radiation, while variations in \( L_\varphi \), \( a \), and \( k \) effectively incorporate the effects of dust. This approach ensures that all three components—dust, radiation, and dark matter—are effectively represented in the dynamics.

\textit{Results}—  
Figure~\ref{fig:k-plots-2} illustrates the evolution of \( \lambda^2 - \kappa^2 \) as a function of \( k \), with fixed parameters and two different choices of \( L_\varphi \). A general trend emerges where the deviation from the bound starts small and increases, reaching a maximum at \( k \approx 0.3 \). The key observation is that, for all cases, BH approach and saturate the chaos bound as the density parameter approaches \( k \approx 1 \). Panel~\ref{fig:k-plots-2}.c highlights that this saturation occurs in the low-temperature limit, as the surface gravity, which corresponds to the Hawking temperature, decreases as \( k \to 1 \).

Figure~\ref{fig:L-plots-2} presents the behavior of chaos in BH as a function of varying parameters: \( L_\varphi \), \( a \), and \( Q \), in panels (a), (b), and (c), respectively. In panel~\ref{fig:L-plots-2}.a, we observe that the chaotic behavior remains almost identical for different values of \( L_\varphi \), especially at higher \( k \) values. The crucial takeaway is that the general trend indicates that chaos approaches the bound as \( k \) increases. For \( k = 0.99 \), the chaos bound is saturated in all phenomenological BH-effective matter systems, regardless of the value of \( L_\varphi \).

Panel~\ref{fig:L-plots-2}.b reveals that the saturation of the chaos bound occurs in BH-matter systems irrespective of the spin parameter \( a \), provided that the density parameter is close to \( k \approx 1 \).

Finally, panel~\ref{fig:L-plots-2}.c confirms that the saturation of the chaos bound is independent of the radiation density, as it is governed solely by the parameter \( k \). In other words, the chaos bound is saturated regardless of the value of \( Q \), with its behavior determined exclusively by the density parameter \( k \). The last three cases further demonstrate the independence of chaos saturation from both dust and radiation components, solidifying the role of \( k \) as the primary factor in governing chaos.

\textit{Discussion}—  
We have shown that a Kiselev rotating BH surrounded by a mixture of radiation, dust, and dark matter halos saturate the chaos bound. This saturation occurs independently of the radiation and dust densities surrounding the black hole, with the density of dark matter playing a crucial role. Specifically, when the dark matter density (\( \rho \sim \frac{k}{r^2} \)) profile follows  \( k \sim 1 \), the chaos bound is saturated.

Another critical point is that this saturation happens in the low-temperature limit. Our numerical results show that as \( k \to 1 \), the surface gravity \( \kappa \) decreases significantly, causing a drastic drop in the Hawking temperature. This is noteworthy, as similar behavior has been observed in the SYK model, where the chaos bound is saturated in the low-temperature limit \cite{Kitaev2015}.

Our model and results offer considerable opportunity to address the chaotic behavior of a BH phenomenologically subjected to effective matter environments. Among these opportunities is the possibility of understanding the saturation of the quantum chaos bound in Kiselev BHs surrounded by a halo of dust, radiation, and dark matter. While interpreting the density parameter \( k \sim 1 \) in terms of astrophysical measurements is challenging, our phenomenological results still may motivate a promising avenue for studying realistic astrophysical BHs to saturate the quantum chaos bound in the low-temperature limit. In other words, our model of a phenomenologically effective BH–matter system can scramble information as efficiently as the laws of physics allow in that regime. Additionally, we have shown that the chaos bound is never violated in rotating BH-effective fluid matter configurations.

Whether or not the model we identified reveals a proof regarding astrophysical BH saturating the quantum chaos bound in the low-temperature limit remains to be seen. What is clear, however, is that it provides a conjectured connection between quantum chaos literature to phenomenologically more realistic BH environments.

\textit{Acknowledgements}—  
RP thanks Jorge G. Russo for profound discussions and support, as well as the Institute of Cosmos Sciences at the University of Barcelona for their hospitality during the course of this work. RP also extends gratitude to Juan Maldacena, Douglas Stanford, and David Berenstein for their valuable comments during the initial stages of the project.

\small
 % This ensures the correct page number is referenced
 % Ensures proper hyperlinks in PDF
 \bibliographystyle{apsrev4-2}
\bibliography{LyapunovExponent}

% \begin{thebibliography}{}
% \input{yourfile.bbl}
% \end{thebibliography}

\end{document}